\documentclass[fleqn,usenatbib]{mnras}

\usepackage[T1]{fontenc}
\usepackage{ae,aecompl}

\usepackage{graphicx}	
\usepackage{amsmath}	
\usepackage{amssymb}	
\usepackage[usenames]{xcolor}
\usepackage{color}
\usepackage[normalem]{ulem}
\usepackage{hyperref}
\usepackage{cleveref}

\title[]{NaCo polarimetric observations of Sz\,91 transitional disk: a remarkable case of dust filtering\thanks{Based on observations made with ESO Telescopes at the Paranal Observatory under programs ID: 098.C-0420 and 099.C-0336}}

\author[K. C. Mauc\'o et al.]{
Karina Mauc\'o$^{1,2}$,\thanks{E-mail: karina.mauco@uv.cl}
Johan Olofsson$^{2,1}$,
Hector Canovas$^{3}$,
Matthias R. Schreiber$^{2,1}$,
\newauthor
Valentin Christiaens$^{4}$,
Amelia Bayo$^{2,1}$,
Alice Zurlo$^{5}$,
Claudio C\'aceres$^{6,1}$,
\newauthor
Christophe Pinte$^{4}$,
Eva Villaver$^{7}$,
Julien H. Girard$^{8}$,
Lucas Cieza$^{5}$, 
\newauthor 
Mat\'ias Montesinos$^{1,2,9}$,
\\
$^{1}$N\'ucleo Milenio Formaci\'on Planetaria - NPF, Universidad de Valpara\'iso, Av. Gran Breta\~na 1111, Valpara\'iso, Chile\\
$^{2}$Instituto  de  F\'isica  y  Astronom\'ia,  Facultad  de  Ciencias,  Universidad de Valpara\'iso, Av. Gran Breta\~na 1111, 5030 Casilla, Valpara\'iso, Chile\\
$^{3}$European Space Astronomy Centre (ESA/ESAC), Operations Department, Villanueva de la Ca\~nada (Madrid), Spain\\
$^{4}$School of Physics and Astronomy, Monash University, Clayton Victoria, 3168, Australia\\
$^{5}$N\'ucleo de Astronom\'ia, Facultad de Ingenier\'ia y Ciencias, Universidad Diego Portales, Av. Ejercito 441, Santiago, Chile\\
$^{6}$Departamento de Ciencias Fisicas, Facultad de Ciencias Exactas, Universidad Andres Bello, Av. Fernandez Concha 700, Las Condes, Santiago, Chile\\
$^{7}$Departamento de F\'isica Te\'orica, Universidad Aut\'onoma de Madrid, Cantoblanco, 28049, Madrid, Spain\\
$^{8}$Space Telescope Science Institute, 3700 San Martin Drive, Baltimore, MD 21218, USA\\
$^{9}$Chinese Academy of Sciences South America Center for Astronomy, National Astronomical Observatories, CAS, 
Beijing 100012, China
}
\date{Accepted XXX. Received YYY; in original form ZZZ}

\pubyear{2019}

\begin{document}
\label{firstpage}
\pagerange{\pageref{firstpage}--\pageref{lastpage}}
\maketitle

\begin{abstract}
We present polarized light observations of the transitional disk around Sz\,91 acquired with VLT/NaCo at $H$ (1.7{\micron}) and $K_s$ (2.2{\micron}) bands. We resolve the disk and detect polarized emission up to $\sim$0.{\arcsec}5 ($\sim$80 au) along with a central cavity at both bands. 
We computed a radiative transfer model that accounts for the main characteristics of the polarized observations. We found that the emission is best explained by small, porous grains distributed in a disk with a $\sim$45 au cavity. Previous ALMA observations have revealed a large sub-mm cavity ($\sim$83 au) and extended gas emission from the innermost (<16 au) regions up to almost 400 au from the star.
Dynamical clearing by multiple low-mass planets arises as the most probable mechanism for the origin of Sz\,91's peculiar structure. 
Using new $L'$ band ADI observations we
can rule out companions more massive than $M_p$ $\geq$ 8 $M_\mathrm{Jup}$ beyond 45 au assuming hot-start models. The disk is clearly asymmetric in polarized light along the minor axis, with the north side brighter than the south side. Differences in position angle between the disk observed at sub-mm wavelengths with ALMA and our NaCo observations were found. This suggests that the disk around Sz\,91 could be highly structured. Higher signal-to-noise near-IR and sub-mm observations are needed to confirm the existence of such structures and to improve the current understanding in the origin of transitional disks.  


\end{abstract}

\begin{keywords}

protoplanetary disks -- stars: individual: Sz\,91 -- stars: variables: TTauri -- techniques: polarimetric

\end{keywords}


\section{Introduction} \label{sec:intro}
One of the main, unresolved questions in planet formation is how the material in the disk grows from (sub-$\mu$m) Interstellar Medium dust to planetesimals. In the past years, there has been a lot of attention on pressure bumps at the edges of cavities, because they can efficiently trap relatively large dust grains, and those are regions in the disks where dust growth is expected to take place \citep[e.g.,][]{pinilla12, pinilla2015c, vandermarel2013, vandermarel2015, macias2018}.
Transitional disks (TD)--disks with sub-mm cavities--are then, prime targets to study pressure bumps, dust traps, grain growth, and their connection to planet formation.

One of the mechanisms proposed to explain cavities on disks is the accumulation of large particles into pressure bumps created by planet-disk interactions, halting their radial drift to the central star and driving grain growth \citep{pinilla12}.
If the planet producing the cavity is massive enough $M_{p} > 5 M_\mathrm{Jup}$, small and large particles are going to get trapped at the inner edge of the cavity; if instead a low-mass planet ($M_{p} \sim 1M_\mathrm{Jup}$) is present, then the small particles will filtrate into the cavity reaching inner regions \citep{pinilla2016a}. This mechanism seems to explain the differences in radii observed not only between the gaseous and the dusty disk, but also between dust grains of different sizes \citep[e.g,][]{dong2012,dejuanOvelar2013,follette2013,garufi2013,vandermarel2016,vandermarel2018}.

Dead zones have also been invoked as an alternative mechanism by which particles accumulate in protoplanetary disks producing ring-like structures. In this case, no planets are needed in order to generate the pressure gradient required for dust grains to get trapped. Instead, low-ionization regions in the disk locally reduce the magnetorotational instability (MRI) which causes the gas flow to significantly decrease, accumulating particles near the boundary of the dead zone \citep{pinilla2016b}.     

High resolution images of protoplanetary disks at sub-mm wavelengths, like the ones reported in the Disk Substructures at High Angular Resolution Project \citep[DSHARP;][]{andrea2018,andrews2018,birnstiel2018,dullemond2018,guzman2018,huang2018a,huang2018b,isella2018,kurtovic2018,zhang2018} have shown with stunning details the variety of sub-structures present in these young systems. Multiple rings and gaps, spiral arms, azimuthal asymmetries/vortices, seem to be common features in planet-hosting disks.

Similar advances have been made at optical/NIR wavelengths thanks to the polarimetric differential imaging technique \citep[PDI; e.g.][]{canovas2011,tsukagoshi2014,benisty2017,avenhaus2018}, which uses the polarized light scattered at the disk surface, to obtain linear Stokes parameters of the incoming light, without being contaminated by the stellar contribution which is mostly unpolarized. Since the scattered polarized light will depend on the properties of dust grains (reflectivity, albedo, porosity, composition), polarimetric observations provide a useful tool to estimate grain properties in protoplanetary disks. 

High-contrast imaging in the IR is a useful technique to detect the companions that could be responsible for the structures observed in protoplanetary disks \citep[e.g.][]{Absil2010, Bowler2016}. Nowadays, adaptive-optics (AO) assisted observations with the so-called angular differential imaging technique \citep[ADI;][]{Marois2006} are routinely used to search for these young giant exoplanets. Using this strategy with several AO-equipped instruments has allowed to obtain the first images of a protoplanet in the large cavity of the disk surrounding T-Tauri star PDS~70 \citep{Keppler2018,Muller2018,Christiaens2019a}.

Sz\,91 is an M0 T Tauri Star (TTS), located in the Lupus III molecular cloud \citep{romero2012}, hosting a TD with the largest mm-dust cavity observed in a low-mass star, and with a significant mass accretion rate ($\rm \dot{M}\sim 10^{-8.8}M_{\odot}yr^{-1}$) \citep{alcala2017}. ALMA observations have revealed a ring-like concentration of mm-sized particles peaking at $\sim$95 au, and gas extended emission from less than 16 au up to almost 400 au \citep{canovas2015a,canovas2016,tsukagoshi2019}. 
PDI observations taken with the Subaru Telescope by \citet{tsukagoshi2014} showed a crescent-like emission peaking closer to the star than the ALMA sub-mm data. They suggested that the observed polarized intensity originates at the inner edge of the transition disk.

In this paper, we report new PDI observations of Sz\,91 obtained in the NIR ($H$ and $K_s$ bands) and new high-contrast ADI observations in the thermal IR ($L'$ band), both obtained with VLT/NaCo. 

\section{Observations and data reduction} \label{sec:obs}

\subsection{$H$ and $K_s$ band Polarimetry}
\label{sec:pol_obs}

We observed Sz\,91 in visitor mode with the NaCo instrument \citep{lenzen2003,rousset2003} at the VLT/UT 1 on March 21st, 2017. The observations were carried out in the polarimetric mode using the broad-band NaCo $H$ and $K_{s}$ filters ($\lambda_c$ = 1.66, 2.18 {\micron}, respectively). In this observing mode a half-wave plate (HWP) first rotates the polarization plane of the incoming light and then a Wollaston prism splits the light into two orthogonally polarized beams, which are projected on different regions of the detector. The pixel size of the camera was set to 0.{\arcsec}027 px$^{-1}$, the readout mode to {\tt Double RdRstRd} and  detector mode to {\tt HighDynamic} ensuring a $\sim$68 e- readout noise and a $\sim$110,000 e- linear dynamic range. As Sz\,91 is relatively faint and red ($m_{\rm 2MASS\ H,\ K_{s}} = 10.1,\ 9.8$) we used the {\tt N20C80} dichroic that sends 80\% of the light to the AO wavefront sensor and 20\% to the detector to maximise the throughput of our observations.

The observations were divided in several polarimetric cycles where each cycle contains four datacubes, one per HWP position angle (at 0{\degr}, 22.5{\degr}, 45{\degr}, and 67.5{\degr}, measured on sky east from north). The airmass ranged from 1.0 to 1.5 during the complete sequence, which included observations of the comparison star (GSPC S264-D). We used detector integration times (DIT) of 15 and 30s for the $H$ and $K_s$ band observations, respectively. During the $K_s$ band observations the seeing was very good and stable with a value of 0.{\arcsec}53$\pm$0.{\arcsec}08. The observing conditions degraded during the $H$ band observations and the average seeing increased to 0.{\arcsec}76 $\pm$ 0.{\arcsec}14. Standard calibrations including darks and flat fields, as well as observations of a photometric standard star (GSPC S264-D) were provided by the ESO observatory. Our observations are summarised in Table~\ref{tab:obs_tab}.

\begin{table}
	\begin{center}
	\caption{Summary of observations}
	\label{tab:obs_tab}
	\begin{tabular}{llcccc} 
		\hline
		\hline

		Date & Band & DIT & NDIT & $t_{\rm exp}$ & $<$seeing$>$\\
        & &(s) & & (s)  & ({\arcsec})\\
        \hline
        2017-03-21 & $H$  & 15  & 4 & 4268$^{\mathrm{a}}$ & 0.{\arcsec}76 $\pm$ 0.{\arcsec}14\\
		2017-03-21 & $K_s$ & 30 & 8 & 8640 & 0.{\arcsec}53 $\pm$ 0.{\arcsec}08\\ 
        2017-04-11 & $L'$  & 0.2 & 100 & 1080 & 0.{\arcsec}93 $\pm$ 0.{\arcsec}11\\
        2017-04-12 & $L'$  & 0.2 & 100 & 2760 & 0.{\arcsec}84 $\pm$ 0.{\arcsec}10\\
        2017-05-15 & $L'$  & 0.2 & 100 & 2280 & 1.{\arcsec}26 $\pm$ 0.{\arcsec}18\\
		\hline
	\end{tabular}
	\end{center}
    \begin{footnotesize}
{\bf Notes.} The fifth column indicates the total (including the four HWP position angles) exposure time for the H and $K_s$ bands.
$^{\mathrm{a}}$We discarded nearly 2/3 of these images due to bad weather.\\
\end{footnotesize}
\end{table}

The two simultaneous, orthogonally polarized images recorded on the detector when the HWP is at 0{\degr}(45{\degr}) were subtracted to produce the Stokes parameter $Q^+(Q^-)$. Repeating this process for the 22.5{\degr}(67.5{\degr}) angles produces the Stokes $U^+(U^-)$ images. The total intensity (Stokes I) was computed by adding all the images. We used customised scripts to process the raw data following the imaging polarimetry pipeline described by \citet{canovas2011}. First, each science frame was dark current subtracted and flat-field corrected. Hot and dead pixels were identified with a $\sigma$-clipping algorithm and masked out using the average of their surrounding good pixels. The two images recorded in each science frame were aligned with an accuracy of 0.05 pixels as described in \citet{canovas2011,canovas2015b}. 
This process was applied to every science frame resulting in a datacube for each Stokes $Q^{\pm}, U^{\pm}$ parameter. The images were median combined and corrected for instrumental polarization using the double-difference method as described in \citet{canovas2011} to produce the final Stokes Q and U images. 

We then derived the polarization angle (P$_{θ}$ = 0.5 arctan (U/Q)), the polarized intensity ($P_I = \sqrt{Q^2 + U^2}$), and the $Q_{\phi}$ and $U_{\phi}$ images following the Stokes formalism \citep[see,][]{schmid2006}:

\begin{equation}
    Q_{\phi} = + Qcos(2\phi) + Usin(2\phi)
    \end{equation}

\begin{equation}
        U_{\phi} = - Qsin(2\phi) + Ucos(2\phi),
\end{equation}

\noindent
where $\phi$ is the position angle of the image coordinates (x, y) with respect to the star location (x$_0$ , y$_0$):

\begin{equation}
    \phi = arctan \frac{x-x_0}{y-y_0} + \theta,
\end{equation}

\noindent
with $\theta$ as the offset needed to correct for instrumental polarization produced by the angular misalignment of the HWP.
This is a convenient coordinate system since, under single scattering assumption, all the emission from a protoplanetary disk should be in the azimuthal direction and be observed as a positive signal in $Q_{\phi}$, whereas emission on $U_{\phi}$ can be taken as disk residual noise \citep{schmid2006}. The central r $<$ 4 px are dominated by noise, and therefore this region has been masked out and is not considered for the analysis.

Many of the individual frames were slightly overexposed and have saturated and/or non-linear pixels around the star projected center. We have median combined a subsample of unsaturated frames to construct a representative point spread function (PSF) for each band. From these PSFs we derive a full width at half maximum (FWHM) of 0.{\arcsec}14 and 0.{\arcsec}19 at $K_s$ and $H$ band, respectively. The narrower FWHM at $K_s$ band is most likely related to a better AO performance, as Sz\,91 is slightly brighter towards redder wavelengths and the weather conditions were more stable during the $K_s$ observations. Combining these two PSFs with the zero points derived from the observations of the standard star we find that the measured flux is consistent, within an error bar of 0.05 mag, with the published 2MASS photometry. We therefore use the 2MASS photometry to calibrate our observations.

At $K_s$ band the disk is clearly detected, while a preliminary analysis of the $H$ band data-set showed that frame selection had to be applied in order to recover the disk signal. We performed data reductions using different subsets of $H$ band observations in order to obtain the highest signal to noise (S/N) disk image. The $H$ band results here presented were obtained after processing a subset with total exposure time of 1440 s and average seeing of 0.{\arcsec}57 $\pm$ 0.{\arcsec}05. 

\subsection{$L'$ band imaging}
\label{sec:Lband_img}

In order to search for (sub)stellar companions, we observed Sz\,91 with NaCo at $L'$ band ($\lambda_c = 3.80 \mu$m) on 11 April, 12 April and 15 May, 2017. 
The first two datasets were obtained in average and relatively stable conditions, while the last one was acquired under mediocre and more variable seeing.
Since Sz\,91 is relatively faint \citep[$m_L \approx 9.7$ mag;][]{Wright2010}, no coronagraph was used.
All observations were obtained in pupil-tracking mode. The DIT was set to 0.2s and data were obtained in cube mode with 100 frames per cube (NDIT = 100). With this choice of DIT, neither the background thermal emission nor the star itself saturated on the detector. The star was jittered in the three good quadrants of the detector throughout the observing sequence, to allow for an optimal sky subtraction.  
We considered the plate scale of NaCo to be 0\farcs0271 $\pm$ 0.0002 px$^{-1}$ in $L'$ band, as per the astrometric calibrations presented in \citet{Milli2017}. Details of the observations can be found in Table~\ref{tab:obs_tab}. 

We implemented our own pipeline to calibrate the data, which is similar to the one used for the NaCo data presented in \citet{Milli2017}. Our pipeline is based on routines of the Vortex Imaging Package \citep[\texttt{VIP};][]{GomezGonzalez2017}\footnote{\url{https://github.com/vortex-exoplanet/VIP}}, an open-source set of python codes for calibration and post-processing of high-contrast images. Our calibration procedure consists of dark subtraction, flat-fielding, bad pixel correction, sky subtraction, centering of the star and bad frames rejection. For the centering, we fitted the stellar PSF with a Moffat function, and shifted frames to place the stellar centroid on the central pixel of all images. 

Given the relatively short integration of each individual dataset, we combined them all in a single datacube. The total parallactic angle rotation achieved in the combined datacube is 111.6\degr.
We then removed frames with the least correlated stellar PSF compared to the median of all PSF images, as measured with the Pearson correlation coefficient. About 10\% of all frames were removed on that basis. This trimming ensured a good PSF modeling and subtraction in post-processing. For the latter, we used principal component analysis coupled with ADI \citep[PCA-ADI;][]{Amara2012,Soummer2012} as implemented in \texttt{VIP}. We considered PCA-ADI either in full frames or in frames divided in 2-FWHM wide concentric annuli. In the latter case, a threshold in parallactic angle corresponding to 1 FWHM azimuthal motion is used to build the PCA library for each annulus \citep[see e.g.][]{Absil2013a}. This is to minimize self-subtraction of any putative companion. 

\section{Modeling and Results} \label{sec:results}

\begin{figure*}
    \centering
    \includegraphics[scale=0.7]{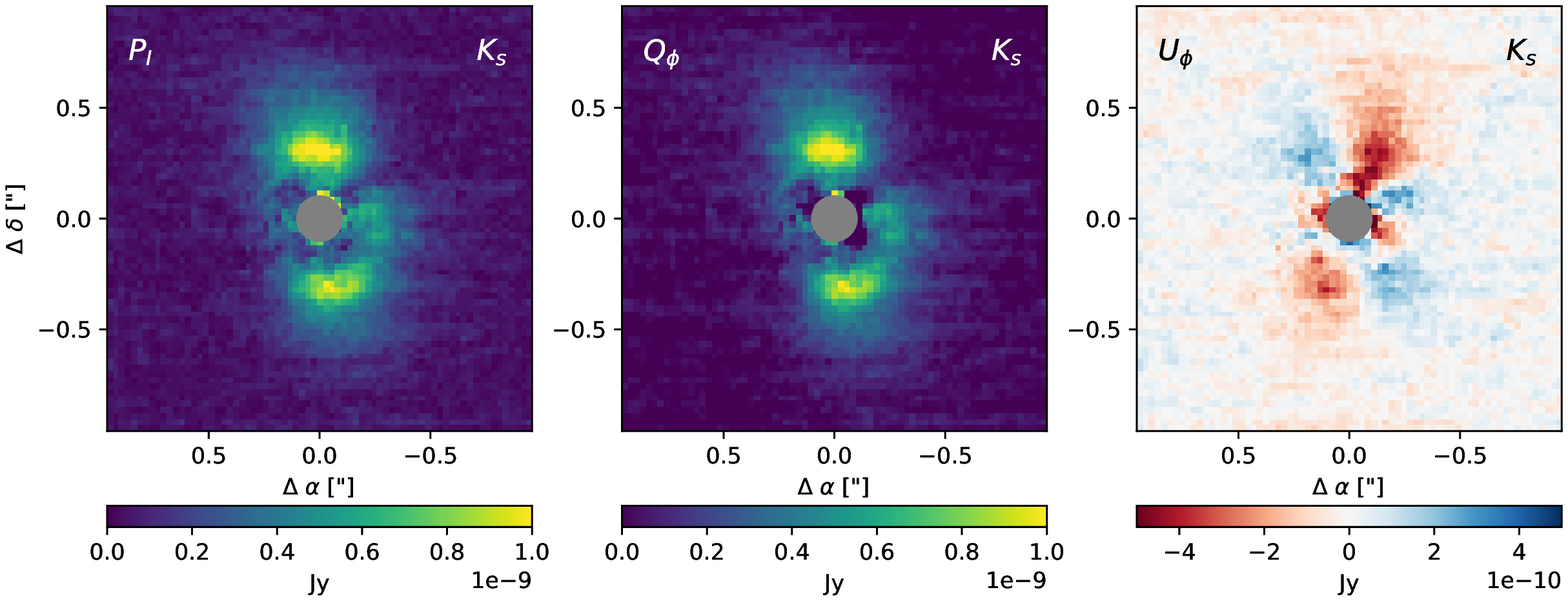}
    \includegraphics[scale=0.7]{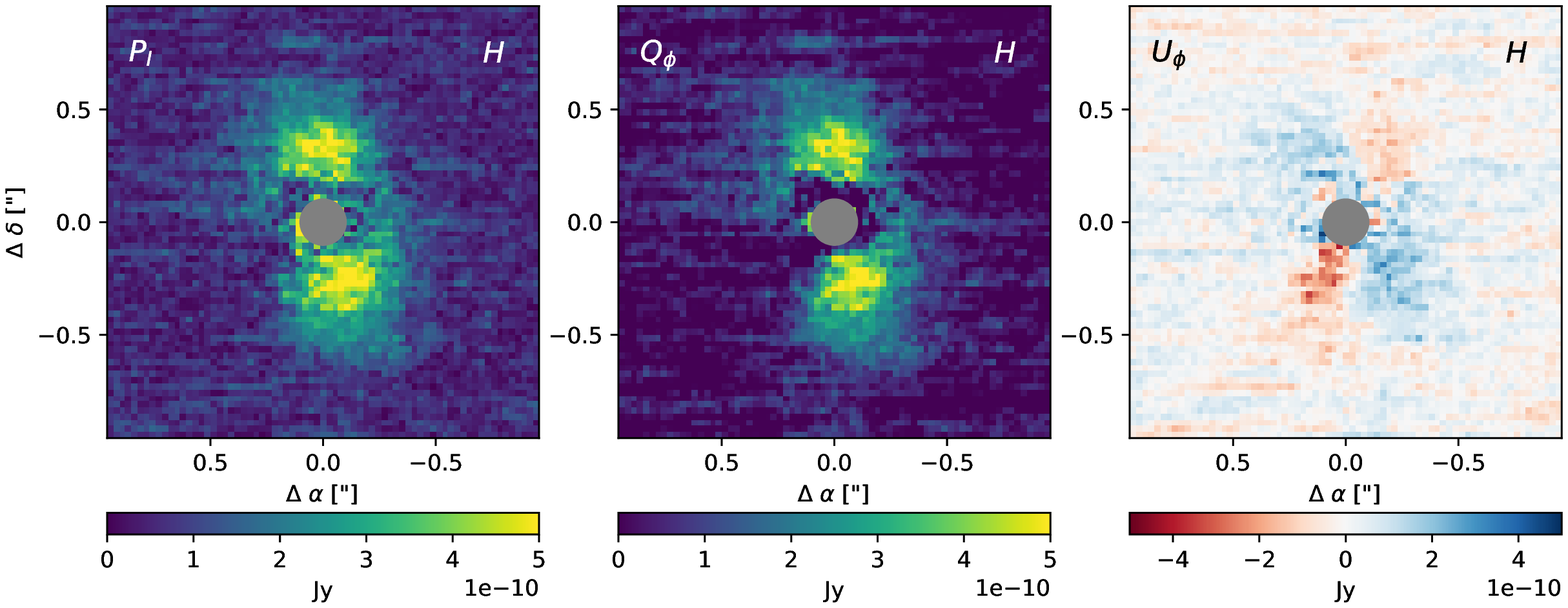}
    \caption{From {\it left to right}: observed polarized intensity, $Q_{\phi}$, and $U_{\phi}$ image at $K_s$ (top) and $H$ (bottom) band. The central 0.{\arcsec}1 region, dominated by noise, has been masked. The observing conditions degraded during the $H$ band observations. In the $U_{\phi}$ images blue corresponds to positive values, red to negative values. North is up and east is to the left in all panels.}
    \label{fig:obs_img}
\end{figure*}

Figure~\ref{fig:obs_img} shows the observed polarized intensity (left), the $Q_{\phi}$ (center), and $U_{\phi}$ (right) images for the $K_s$ (top) and $H$ (bottom) bands. 
We detect two lobes north and south of the star at both bands which correspond to the major axis of the disk. Fainter emission is also seen on the minor axis at the right side, which has been identified as the front-facing side of the disk closest to us \citep{tsukagoshi2014,tsukagoshi2019}. The disk shows polarized emission above noise level up to $\sim$0.{\arcsec}52 along the major axis in the $Q_{\phi}$ maps. A central cavity is observed at both bands as seen by the substantial emission dips close to the center. The residual signal observed in the $U_{\phi}$ images, especially at $K_{s}$ band, might be related to multiple scattering events where the linear polarization is not purely azimuthally but have a radial contribution, an effect that is even more pronounced for disks with inclinations $\geq$ 40{\degr} \citep{canovas2015c,pohl2017}. 

In this section, we aim to provide a radiative transfer model for
Sz\,91, that reproduces the main characteristics of the polarized emission observed at $H$ and $K_s$ bands. We first estimate the stellar parameters of the source (model inputs) and then fit a radiative transfer model to the observations. 

\subsection{Stellar Properties: VOSA} \label{sec:stellarprop}
We estimated the stellar properties of Sz\,91 using the Virtual Observatory (VO), VO-tool VOSA\footnote{\url{http://svo2.cab.inta-csic.es/theory/vosa/}} \citep[Virtual Observatory SED Analyzer;][]{bayo2008}. The observed (stellar) SED of the source is compared to the synthetic photometry obtained using a suit of theoretical models via a $\chi^2$ test. In our case we considered the BT-Settl-CIFIST and Kurucz models in the analysis. For this, we used a distance to the source of $d$ = 159.06$\pm$1.63 pc \citep[Gaia DR2;][]{bailer-jones2018}, and consider the extinction in the line of sight, $A_{\rm v}$, also as a fit parameter with an initial upper limit of 2.5 mag taken from the extinction maps of the IRSA Infrared Science Archive \citep{schlegel1998}. 

The best fit was found for the BT-Settl-CIFIST models which are used to infer the total observed flux from the star. We highlight that this estimate is more accurate than the one obtained using a bolometric correction derived only from a single colour. Then, we locate the object in a Hertzsprung-Russell diagram, given its estimated luminosity, $L_{*}$, and effective temperature, $T_{\rm eff}$, and use the isochrones and evolutionary tracks from \citet{baraffe2015} to estimate the stellar mass, $M_*$, and age of the object. The uncertainties are estimated through a Bayesian approach as explained in Section~\ref{sec:best_fit}.
The stellar radius, $R_{*}$, on the other hand, is estimated using the dilution factor defined as $M_d = (R_{*}/d)^2$, with an uncertainty set by error propagation.  
Table~\ref{tab:stellar_para} lists the stellar parameters obtained in this work along with their uncertainties. 

\begin{table}
	\begin{center}
	\caption{Stellar parameters}
	\label{tab:stellar_para}
	\begin{tabular}{lrr} 
		\hline
		\hline
		Parameter & Value & Uncertainty\\
		\hline
        $T_{\rm eff}$ (K) & 3800  & [3750, 3850]\\ 
        Log $L_{*}$ ($L_{\odot}$) & -0.59 & [-0.63, -0.56] \\
        $R_{*}$ ($R_{\odot}$) & 1.18    &  [1.16, 1.18] \\
        $M_{*}$ ($M_{\odot}$) & 0.58  &  [0.51, 0.62] \\
        Age (Myr)             & 5.0     &  [3.6, 7.4] \\
        $A_{\rm v}$ (mag)         & 1.65    &  [1.58 1.72]\\
		\hline
	\end{tabular}
	\end{center}
\end{table}

The new distance reported in the Gaia DR2 catalog (159 pc; before the source was thought to be located at 200 pc) results in a significantly older age for Sz\,91. Using the above methodology, we obtained an age of $5^{+2.4}_{-1.4}$ Myr, older than the $\sim$3 Myr reported by \citet{tsukagoshi2019}. Since the estimate of ages of individual objects is model dependent and very uncertain, we considered Sz\,91 to be older than at least 3 Myr.  

\subsection{Radiative transfer modeling}

\subsubsection{MCFOST Model}\label{model}

We used 3D radiative transfer code MCFOST \citep{pinte2006,pinte2009} to model the polarimetric images at $H$ and $K_s$ band. MCFOST computes the dust temperature structure and scattering source function, under the assumption of radiative equilibrium between the dust and the local radiation field, via a Monte Carlo method. Images are then obtained via a ray-tracing method, which calculates the output intensities by integrating formally the source function estimated by the Monte Carlo calculations. Full calculations of the polarization are included using the Stokes formalism\footnote{\url{http://ipag-old.osug.fr/~pintec/mcfost/docs/html/overview.html}}. 

The surface density distribution of the disk is described by a simple profile of the form: 

\begin{equation}
    \Sigma(r)\ = \Sigma_{\rm o} \left(\frac{r}{[\mathrm{au}]}\right)^{\gamma},  
\end{equation}
\noindent
where $\Sigma_o$ depends on the mass and size of the disk, and $\gamma$ represents the power-law index of the surface density profile. 
A Gaussian profile is used to describe the vertical density distribution with a disk aspect ratio which is radially parametrized as $H(r) = H_{100} (r/100\,\mathrm{au})^{\psi}$, where $H_{100}$ is the scale height at $r = 100$\,au, and $\psi$ is the flaring index of the disk. In Table~\ref{tab:mod_par} (top) we show the fix model parameters used in this work. We adopted the same $H_{100}$, $\gamma$ and $\psi$ values of \citet{canovas2015a}. 
The optical depth is changed using different disk's dust masses.

We consider dust grains to be irregular in shape by assuming a distribution of hollow spheres (DHS) as our grain type with a maximum volume void fraction of 0.8 \citep{min2005}. We used an inclination of 49.7{\degr} and a position angle (PA) of 18.1{\degr} derived from ALMA observations \citep{tsukagoshi2019} assuming that the polarized emission comes from a region co-planar to the sub-mm ring.
We stress that there is a small degeneracy between the PA and the grain size (i.e. the phase function), therefore in order to sample in more detail the dust properties (grain size, porosity, dust mass) we fixed the PA of the models to the value estimated from the ALMA observations. 
\begin{table}
	\begin{center}
	\caption{Model parameters}
	\label{tab:mod_par}
	\begin{tabular}{lr} 
		\hline
		\hline
		Parameter &  Value\\
		\hline
        $H_{100}$ (au) & 5 \\
        $\gamma$  & -1  \\ 
        $\psi$    & 1.15 \\
        $R_{\rm out}$ (au) & 150 \\
        \hline
        Parameter space \\
        \hline
        $R_{\rm in}$ (au) & 35-55, steps = 5\\
        Porosity  & 0.1-0.9, steps = 0.1\\
        $a_{\rm min}$ ({\micron}) & 0.05-0.175, steps = 0.025\\
        $\delta s$ ({\micron}) & 0.05-0.25, steps = 0.05\\
        $m_{\rm dust}$ ($M_{\odot}$) & $10^{-6}-10^{-7}$, steps = 1 (in log scale)\\
        \hline
	\end{tabular}
	\end{center}
	\begin{footnotesize}
    {\bf Notes.} $a_{\rm max} = a_{\rm min} + \delta s$.\\
    \end{footnotesize}
\end{table}

Once the surface density and temperature structure is computed, synthetic ray-traced polarized images (Stokes I, Q, and U maps) can be produced at any wavelength. To compare with our observations, these images were projected into a grid with pixel size of 0.{\arcsec}027 px$^{-1}$ (equal to the scale of the NaCo/VLT images). Then, they were scaled using the stellar $H$ and $K_s$ 2MASS magnitudes and were convolved using a Gaussian point spread function (PSF) of 2.5-px width size. Finally, we computed monochromatic Stokes $Q_{\phi}$ and $U_{\phi}$ images, at 1.7 and 2.2 {\micron} following the same strategy as for the observations (section~\ref{sec:pol_obs}).

\subsubsection{Best fit} \label{sec:best_fit}

We ran a grid of 13500 models varying the following parameters: the minimum/maximum grain size ($a_{\rm min}$, $a_{\rm max}$), the grain porosity, the size of the cavity ($R_{\rm in}$)  and the dust mass ($m_{\rm dust}$). We fixed the outer radius ($R_{\rm out}$) of the disk to 150 au since it does not affect the final image (it only depends on $R_{\rm in}$). We considered pure silicate grains with a small amount of carbonaceous particles using the dust opacity from \citet{drainelee84}. Table~\ref{tab:mod_par} (bottom) shows the parameter space used in this work. 

We determined both the best fit model as well as the uncertainties using the Bayesian approach. For this, we constructed the probability distribution functions (PDFs) for our model parameters following a Bayesian analysis as described in \citet[][VOSA 6.0]{bayo2008}\footnote{\url{http://svo2.cab.inta-csic.es/theory/vosa/index.php}}, where for each model we assign a relative probability as:
\begin{equation}
    W_i = {\rm exp} (-\chi_i^2/2)
\end{equation}

\noindent
where the subscript $i$ represents each individual model on the grid, and the $\chi_i^2$ represents the goodness of the fit estimated as:

\begin{equation}
\chi_i^2 = \sum_n \frac{(Q_{\phi}^{\rm mod} - Q_{\phi}^{\rm obs})^2}{\sigma^2},
\end{equation}

\noindent
with $n$ the number of pixels included in the fit and $\sigma$ as the standard deviation measured in concentric annulii from the center of the $U_{\phi}^{\rm obs}$ image excluding the central 0.{\arcsec}1 region. For the $\chi^2$ values, the central r $<$ 0.{\arcsec}15 region (dominated by noise) as well as the outer r $>$ 0.{\arcsec}63 region (free of disk emission) of each image have been masked out and were not considered for the analysis. 
Then the probability corresponding to a given parameter value $\alpha_j$ is given by:
\begin{equation}
    P(\alpha_j) = \sum_i W_i
\end{equation}

 The final normalized PDF for each parameter is obtained by dividing by the total probability (the sum of the probabilities obtained for each value): 
 \begin{equation}
     P{\arcmin}(\alpha_j) = \frac{\alpha_j}{\sum_i P(\alpha_i)}
 \end{equation}

\begin{figure*}
    \centering
    \includegraphics[scale=0.7]{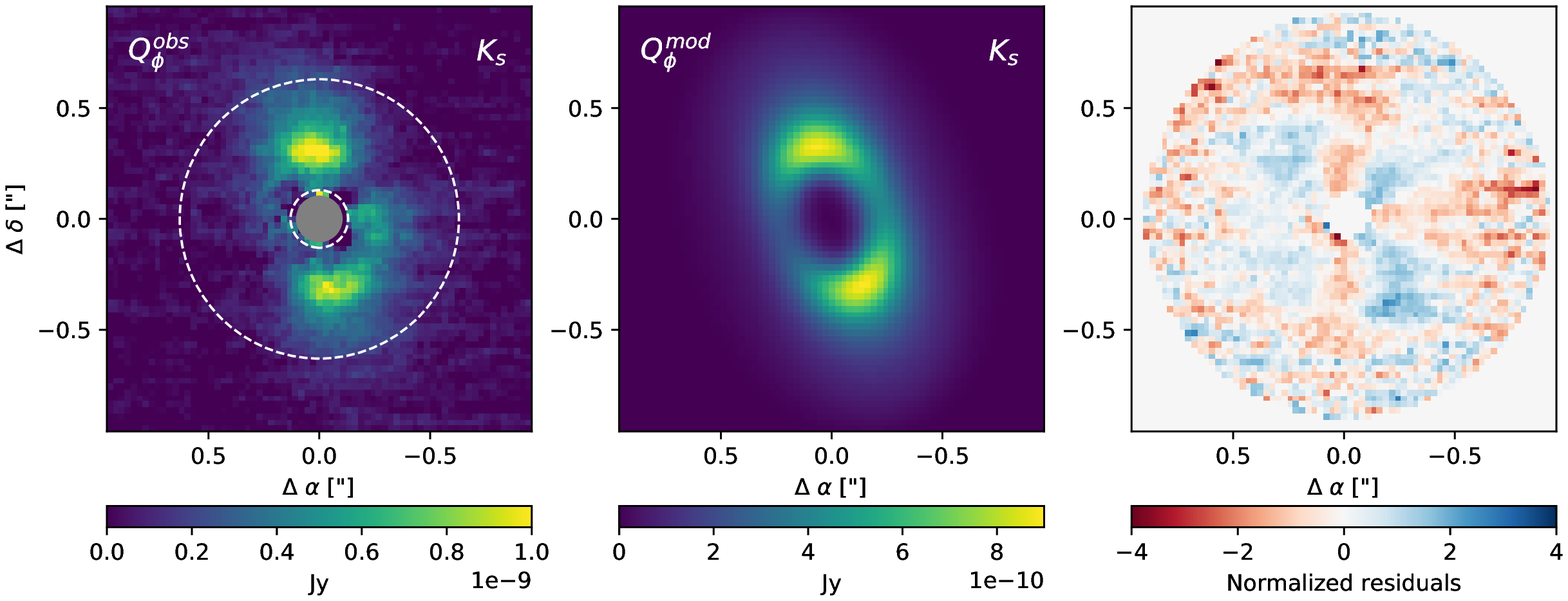}
    \includegraphics[scale=0.7]{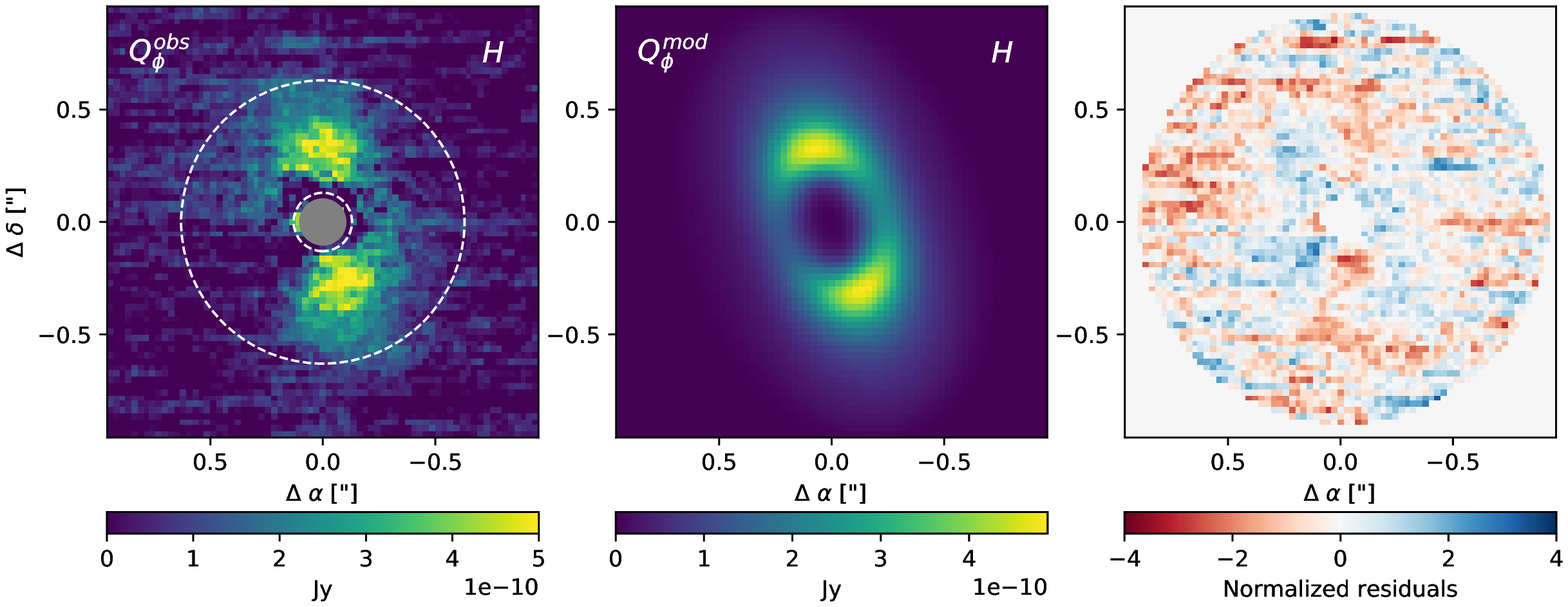}
    \caption{From {\it left to right}: $Q_{\phi}$ observed image, best MCFOST model, normalized residuals for the $K_s$ (top) and $H$ (bottom) bands. The dashed circles plotted in the left panels circumscribe the area taken into account for the $\chi^{2}$ fit. The faint butterfly pattern observed in the residual map at the $K_s$ band can be the result of the structure of the noise in the $U_{\phi}$ image (see text for details). In the right panels blue corresponds to positive values, red to negative values. North is up and east is to the left in all panels.
    }
    \label{fig:best_fit}
\end{figure*}

\begin{figure*}
    \centering
    \includegraphics[scale=0.35]{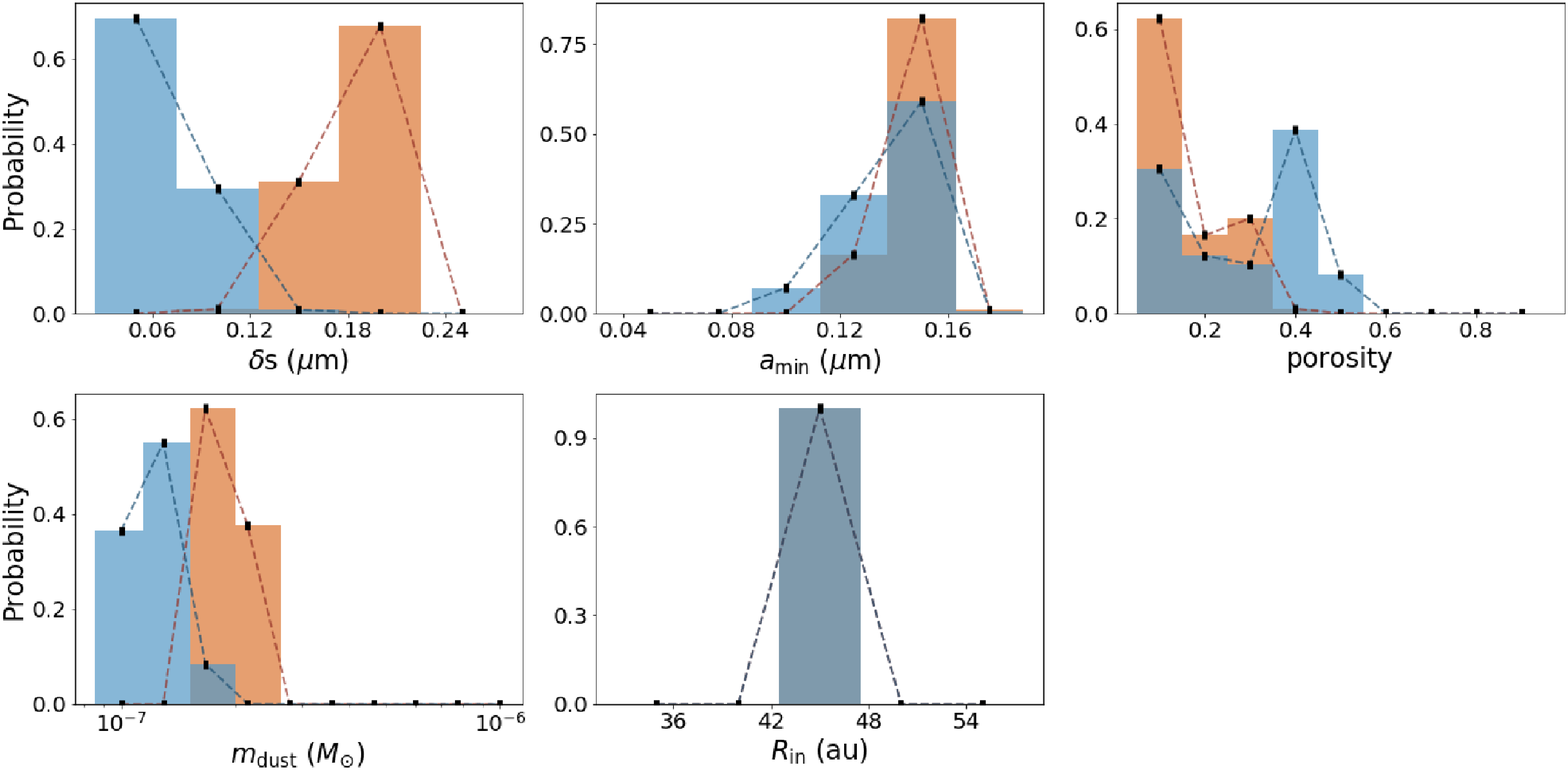}
    \caption{Probability distribution functions (PDFs) of our model parameters at $K_s$ (orange) and $H$ (blue) bands. The confidence intervals reported on Table~\ref{tab:best_val} are estimated for a 68\% (1$\sigma$) confidence level.
    }
    \label{fig:KH_pdfs}
\end{figure*}

Figure~\ref{fig:best_fit} shows the models that best fitted our observations along with their corresponding residuals, which are estimated as $(Q_{\phi}^{\rm mod} - Q_{\phi}^{\rm obs})/\sigma$. The dashed circles plotted in the left panel circumscribe the area taken into account for the $\chi^2$ fit, basically, all the emission coming from the disk. Table~\ref{tab:best_val} lists the estimated parameters along with their uncertainties, which have been defined as the limits encompassing 68\% (1$\sigma$) of the total area around the PDF maximum for each parameter. For those cases where the best parameter falls on one of the edges of the range of values used in the models, we have considered these values as upper or lower limits, and they are indicated by parentheses instead of square brackets in Table~\ref{tab:best_val}.
The PDFs of our model parameters are shown on Figure~\ref{fig:KH_pdfs}. As seen on the Figure, the polarized emission can be explained by small grains ($<$0.4 {\micron}), with moderate porosity ($<$40\%), distributed in a ring  located at $\sim$45 au from the central star.  

\begin{table}
	\begin{center}
	\caption{Best fit values}
	\label{tab:best_val}
	\begin{tabular}{lrrrr} 
		\hline
		\hline
		Parameter & H & uncertainty & K & uncertainty \\
        \hline
        $R_{\rm in}$ (au) & 45 & [43, 47] & 45 & [43, 47]\\
        Porosity     & 0.40 & (<0.40) & 0.10 & (0.1, 0.17] \\
        $a_{\rm min}$ ({\micron}) & 0.15 & [0.12, 0.16] & 0.15 & [0.13, 0.16] \\
        $\delta_s$ ({\micron}) & 0.05 & (0.05, 0.08] & 0.20 & [0.15, 0.22]\\
        $m_{\rm dust}$ ($10^{-7}M_{\odot}$) & 1.29 & [1.07, 1.35] & 1.67 & [1.59, 2.27] \\
        \hline
	\end{tabular}
	\end{center}
	\begin{footnotesize}
    {\bf Notes.} Upper/lower limits are indicated by parentheses instead of square brackets.\\
    \end{footnotesize}
\end{table}

We note, however, that since $\sigma$ is estimated in concentric rings from the center, the stronger signal towards the center of the $U_{\phi}$ image will translate into larger uncertainties at the inner regions and hence, will give preference to models that best match outer regions. This explains the faint butterfly remnant observed in the residual map at $K_s$ band and why we do not observe any feature on the $H$ band, where the $U_{\phi}$ signal is weaker. In any case, we do not see any significant emission on our residual maps besides the noise induced by the faint $U_{\phi}$ signal structure, which reinforces the validity of our modeling. 
 
\subsubsection{Caveats of the modeling}

Reproducing the shape of the polarized (or scattered light) phase function is a known challenge when modeling young disks or debris disks \citep[e.g.,][]{Milli2017}. As discussed in \citet{min2016}, the polarized phase function may best trace the optical properties of the smallest constituents of dust grains, remaining insensitive to large-scale structures such as aggregates. For this reason, the grain size distribution inferred from our modeling results may be biased towards smaller sizes. As noted above, the models that best explain our NaCo observations are those having a very narrow range of grain sizes (Table~\ref{tab:best_val}); similar results were also found for the debris disk around HD\,61005 \citep{olofsson2016}. Grains may well be in the form of aggregates, and the polarized observations would then be dominated by the small monomers (see Sec.~\ref{sec:dust_prop_pol}). Our analysis, based on the observational data currently at hand, suggests that small dust grains are indeed present at the disk surface layers of Sz\,91, however, in what shape or form (and as a consequence the exact grain size distribution) remain uncertain. Therefore, both the total dust mass and the optical depth reported here should be treated carefully. 

\subsection{Companion detection limits} \label{sec:Lband_limits}

Our final $L'$ PCA-ADI images did not reveal any significant point source, for a wide range of tested number of principal components (between 1 and 100).
We used \texttt{VIP} (Sect.~\ref{sec:Lband_img}) to compute the $5\sigma$-contrast curve achieved by annular PCA-ADI, using the number of principal components that optimizes contrast at each radial separation. 

\begin{figure}
    \centering
    \includegraphics[scale=0.40]{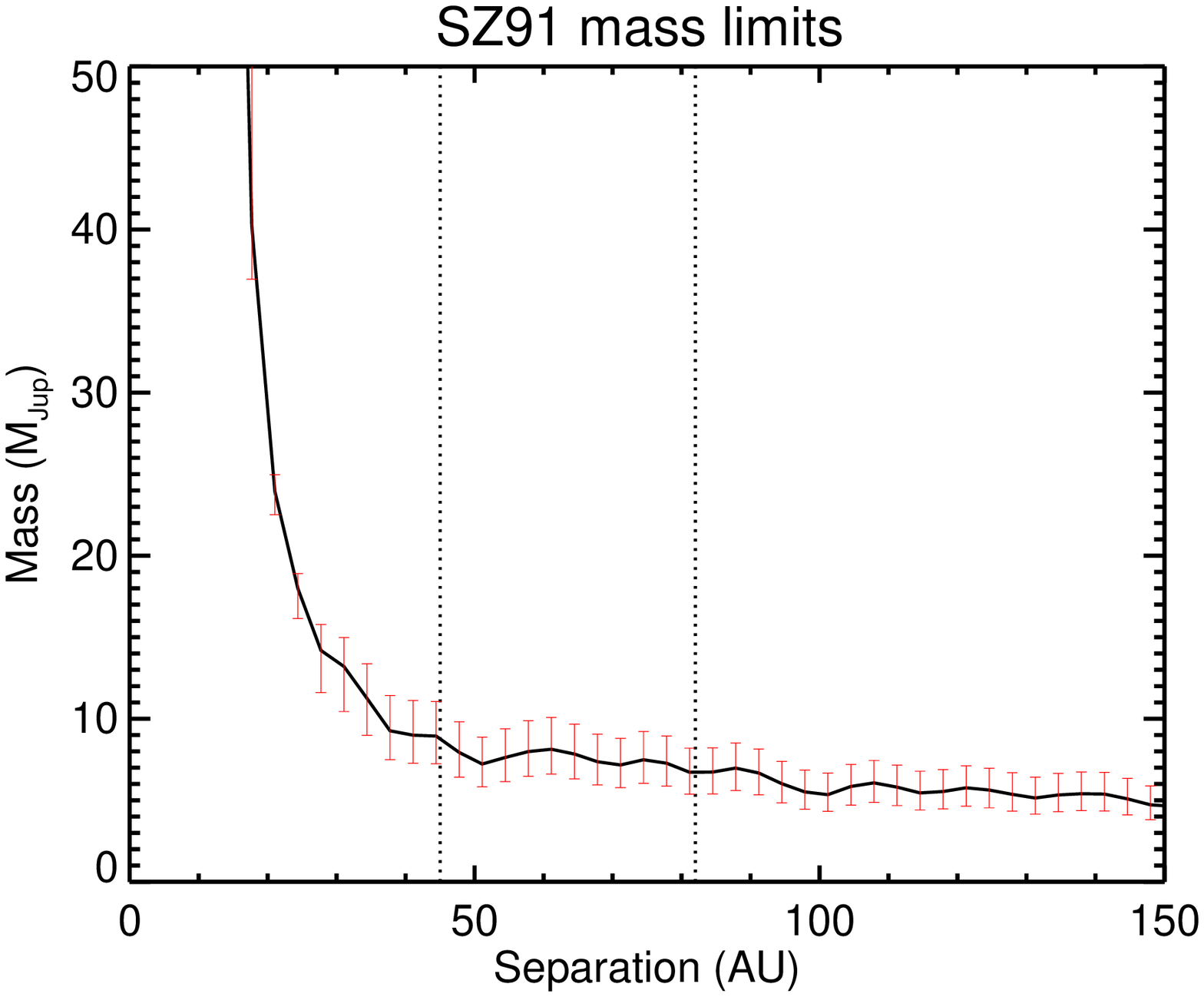}
    \caption{Contrast curve from the NaCo $L'$ band observations derived using the COND/Dusty models \citep{allard2001} for an age of 5 Myr in Jupiter masses (solid line). Error bars indicate the Sz\,91 age uncertainty (Table~\ref{tab:stellar_para}). Dotted lines indicate the location of the dust cavities radius observed in polarized light at 45 au (from this work) and in the sub-mm at 82 au (from \citet{tsukagoshi2019}). We can rule out massive giant planets ($\geq$8 $M_\mathrm{Jup}$) orbiting beyond 45 au.
    }
    \label{fig:Lband_masslimit}
\end{figure}

We then used the COND/Dusty models for brown dwarfs and giant planets atmospheres \citep{allard2001} to convert the contrast curve into mass sensitivity limits. We used an age of 5 Myr. Giant planets with masses
above $\sim$8 $M_\mathrm{Jup}$ orbiting beyond 45au could be detected in our observations, as shown on Figure~\ref{fig:Lband_masslimit}. Note that inside the innermost (r < 45 au) regions, massive giant planets can still be present.
We remark, however, that the uncertainties are probably very high given the sensitivity of the planet brightness to the initial conditions of planet formation \citep{mordasini2013}, which in our case are based on ``hot start models", and that the disk emission may mask the planet signal. 

\section{Discussion} \label{sec:Disc}

In this section, we inquire about the potential origin of the observed cavity around Sz\,91 based on our NaCo data and ALMA observations from previous studies. We also discuss the implications of asymmetric features observed in the polarized emission and in the phase function profile. 

\subsection{A large cavity in the disk around Sz\,91; evidence for dust trapping}


ALMA data of Sz\,91 has revealed a sub-mm narrow ring at $\sim$95 au from the central star, along with $^{12}$CO (3-2) emission extending from the innermost regions (< 16au) up to almost 400 au \citep{tsukagoshi2019}. 
Previous modeling of the $^{12}$CO (3-2) emission made by \citet{vandermarel2018}, showed a gas-depleted cavity at 39.7 au (after scaling with the new Gaia distance) from the star. This apparent inconsistency may be a sensitivity issue, since the \citet{tsukagoshi2019} observations are about twice more sensitive. Additionally, \citet{vandermarel2018} used a simplified model of sharp gas cavity edges, whereas in reality these edges are probably not sharp but rather smooth transitions. This could potentially account for the differences observed by these authors. 
One should note, however, that even when the gas reaches at least 16 au according the models of \citet{tsukagoshi2019}, estimated based on blueshifted
emission near the highest velocity on the $^{12}$CO channel map, their $^{12}$CO and HCO$^+$ moment maps also suggest that there is a lack of signal in the inner regions (rapidly decrease of emission towards the star in their Figures 4 and 6, although within one beam of resolution). This could be in line with a drop of density inwards of 40 au (a gas-depleted cavity), as suggested by \citet{vandermarel2018}. Nevertheless, this should be treated carefully due to the different angular resolution and sensitivity of these observations.

\begin{figure}
    \includegraphics[scale=0.40]{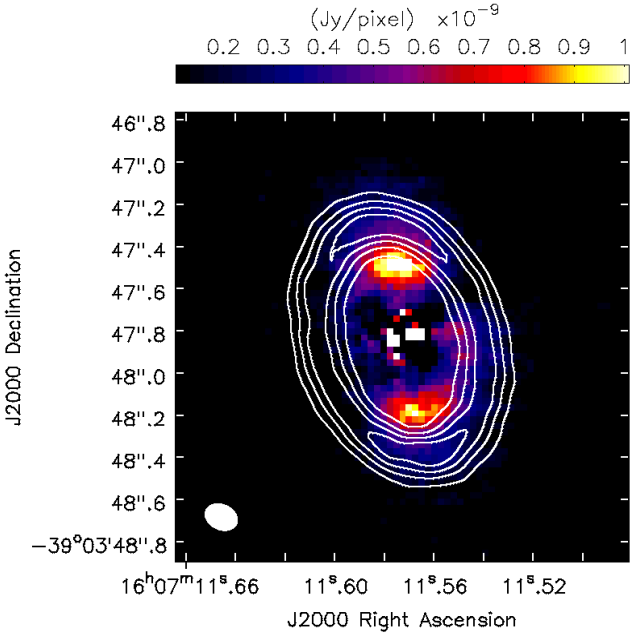}
    \caption{ALMA Band 7 continuum archival image of Sz\,91 from project ID 2015.1.01301.S (white contours) overlaid on the $K_s$ band $Q_{\phi}$ NaCo image (color scale). The ALMA synthesized beam is shown as the filled ellipse at the bottom-left corner of the plot. The polarized emission observed with NaCo lies inside the sub-mm cavity. 
    }
    \label{fig:ALMA_NACO}
\end{figure}

Polarimetric data, on the other hand, showed a ring-like structure of small (<0.4{\micron}) grains peaking inside the sub-mm cavity as seen in Figure~\ref{fig:ALMA_NACO}, where we show the ALMA Band 7 continuum archival image of Sz\,91 from project ID 2015.1.01301.S (white contours), overlaid on the $K_s$ band $Q_{\phi}$ NaCo image (color scale).
TD showing different radii between the gas and dust material, particularly a larger gas extent than (sub)mm dust, have been observed in the past \citep[e.g,][]{panic2009,andrews2012,rosenfeld2013,canovas2016,ansdell2018,gabellini2019}. 

A few physical mechanisms have been proposed to explain the formation of a large (sub)mm cavity:
dynamical clearing  by stellar or planetary companions \citep[e.g.,][]{zhu2011,Pinilla2012b}, an extended dead zone \citep[e.g.,][]{flock15,pinilla2016b}, and internal photoevaporation due to irradiation from the central star \citep{alexander_armitage07}.
Models of internal photoevaporation predict 
dust cavities smaller than $\sim$20 au
and accretion rates of less than $\rm 10^{-9}M_{\odot}yr^{-1}$
\citep{owen2011, ercolano2017}. Therefore, the presence of such a large sub-mm cavity around Sz\,91, along with its relatively high mass accretion rate of $\rm \dot{M}\sim 10^{-8.8}M_{\odot}yr^{-1}$ \citep{alcala2017}, make this mechanism very unlikely.

\subsubsection{Dynamical clearing by stellar or planetary companions}

We have analyzed one epoch of high-resolution Las Campanas/MIKE data obtained on June 2014 using the 1.0{\arcsec} slit. We focus on data
from the red arm of MIKE which covers 4900 to 9500 {\AA} with a S/N $\sim$10. The data were reduced using the MIKE pipeline in the
Carnegie Observatories' CarPy package \citep{kelson2000,kelson03}. From this single epoch we can discard an SB2 nature of the object, therefore equal mass binaries and mass ratios above 0.7 can be discarded out to 1au.
Furthermore, \citet{romero2012} excluded a stellar companion down to separations of $\sim$30 au and \citet{melo2003} found no evidence for a close-in binary companion in their 3 yr radial velocity survey down to masses of $\approx$ 0.2 $M_{\odot}$. 

This is in agreement with the work of \citet{villenave2019} where they found that Sz\,91's possible companions must be in the planetary mass regime ($M_{\rm p}< 13 M_\mathrm{Jup}$), according to the prescription of \citet{dejuanOvelar2013} which relates the ratio between the radius of the scattered light cavity and that of the sub-mm ring. We remark, however, that using the updated radius for the peak of the sub-mm ring from \citet{tsukagoshi2019} and that of the scattered light cavity from this work ($R_{\rm in}$ on Table~\ref{tab:mod_par}), and within the uncertainties, Sz\,91 seems to fall in the region of companion masses above 13 $M_\mathrm{Jup}$ but it is very close to the limit between companion masses above or below this value (red shaded region in Figure 10 of \citealp{villenave2019}). According to our new $L'$ band ADI observations (Sect.~\ref{sec:Lband_limits}), on the other hand, we found a mass sensitivity limit for any putative planet of $M_{\rm p}$ $\leq$ 8 $M_\mathrm{Jup}$ beyond 45 au. Within $35$\,au, our sensitivity constraints are poor and we cannot rule out the presence of brown dwarf companions (Figure~\ref{fig:Lband_masslimit}).
Follow-up observations are needed in order to completely rule out a stellar companion, however, we will mainly focus on companions of planetary origin hereafter.

In the planet scenario, a large sub-mm cavity is created along with a pressure bump at the outer edge of the cavity.
Small dust coupled to the gas move at sub-Keplerian velocities while large particles ($>$ 1mm) move with a Keplerian motion. This difference in velocity causes big grains to experience a head wind driven by the gas movement. This makes large particles to loose angular momentum and fall into inner radius. If a positive pressure gradient exists, as a consequence of a planet carving a cavity in the disk, then these particles will get trapped into pressure bumps located at the outer edge of the newly formed cavity. 

The fact that the emission from small ($\mu$m) grains, as probed by our NaCo observations, peaks inside the sub-mm cavity, suggests partial filtration of dust. This could happen in the presence of {\it low-mass} planets as small grains may not be completely filtered at the outer edge of the planet induced-cavity, and hence pass through the edge to inner regions.
Even though \citet{tsukagoshi2014} suggested that the polarized intensity emission at $ K_s$ band was the result of light coming from the inner edge of the disk, it might very well be that this emission is caused by small, optically thin dust passing through the pressure bump into inner regions. Our modeling reveals optical depths around 0.2-0.3, so the polarized emission is rather partially optically thin. In fact, lower optical depths are expected since our models only used small grains.
Optically thin dust inside cavities of several TD and pre-transitional disks have been observed \citep[e.g.,][]{calvet05,espaillat07, espaillat10,follette2013,mauco2018,perez_alice_2018}. 

With respect to the gas distribution, if the gas is not depleted inside the sub-mm cavity and embedded planets are indeed the cause of this structure, then these planets must be of low-mass \citep[0.1 - 1 $M_{\rm Jup}$,][]{Pinilla2012b,zhu12,rosotti2016}. 
This is also consistent in the case of a gas-depleted cavity, as suggested by \citet{vandermarel2018}, since multiple low-mass planets can lead to shallower cavities with depletion factors of at least an order of magnitude \citep{duffell2015}. 
Embedded giant planets, on the contrary, have shown to produce deeper cavities in both, the gas and the dust component on protoplanetary disks \citep[e.g,][]{rice2006,pinilla2016a, pinilla2016b,gabellini2019}.

\subsubsection{An extended dead zone}

Dead zones have also been invoked to explain TD structures. These are low-ionization regions on the disk where the high energy (X-rays and UV) radiation from the star cannot penetrate and, as a consequence, the MRI is suppressed.
At the outer edge of these low ionization regions 
a bump in the gas density profile is created, due to
the change of accretion from the dead to the active MRI zones. Strong accumulation of (sub)mm-sized particles are expected at the location of the outer edge of the dead zone, while the gas is only slightly depleted in the inner part of the disk \citep[e.g,][]{flock15, pinilla2016b}, as seem to be the case of Sz\,91 according to \citet{tsukagoshi2019}.
If we considered, on the other hand, a gas-depleted cavity \citep{vandermarel2018}, then it requires the inclusion of a MHD wind to the dead zone in order to create the spatial segregation between the distribution of gas and dust \citep{pinilla2016b,pinilla2018}.
In this case, the gas surface density inside the cavity can be depleted by several orders of magnitude and increases smoothly with radius, in agreement with the $^{12}$CO and HCO$^+$ moment maps of \citet{tsukagoshi2019}.
Nonetheless, dead zones always produce a highly depleted gaseous outer disk, which is not the case of Sz\,91 ($^{12}$CO is observed up to $\sim$400 au).

Additionally, \citet{pinilla2016b} studied the effects of a large dead zone (with an outer edge at $\sim$40 au) in the radial evolution of gas and dust in protoplanetary disks through MHD simulations. 
On their polarized synthetic images they observed that small (0.65{\micron}) grains lie just in front of the pressure maximum and slightly closer to the central star than large (mm) grains, although this segregation is very small. In fact, they concluded that this scenario always produces dust cavities
at short {\it and} long wavelengths of similar size at the location of the pressure bump, contrary to what is found in Sz\,91 (i.e. $\mu$m-sized particles closer in than sub-mm grains).

\subsubsection{Final remarks}
Overall, Sz\,91 morphology suggests that whatever is the origin of the sub-mm cavity allows enough gas to reside in the inner regions of the disk.
One way to discriminate between these gap opening mechanisms is to radially resolve the sub-mm ring. Models of embedded planets predict a radially asymmetric ring with a wider outer tail at early times, while dead zones always produce radially symmetric ring-like structures \citep{pinilla2018}. However, for a $\sim$5 Myr old star it is not clear if this diagnostic still applies. 
Higher angular resolution observations are needed
to validate whether or not low-mass embedded planets are the most likely mechanism for the origin of the cavity in Sz\,91.

\subsection{Apparent ``dip" on the polarized emission} 
\begin{figure*}
    \centering
    \includegraphics[scale=0.7]{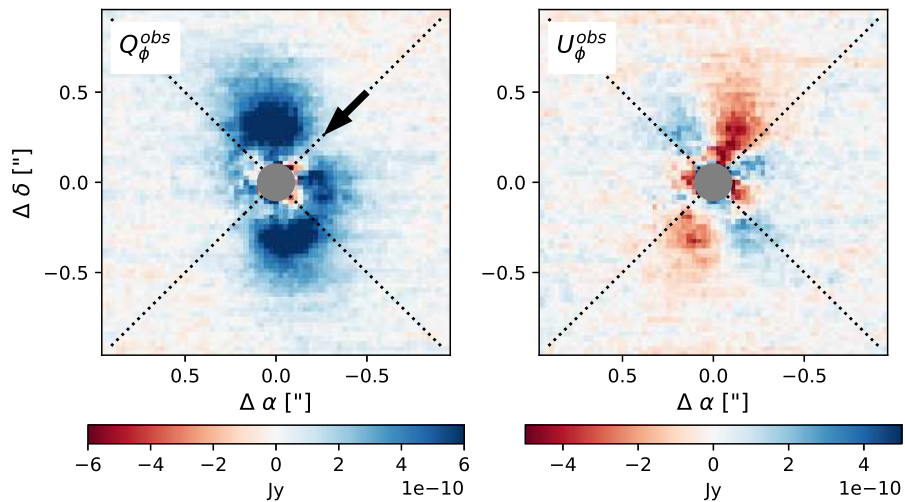}
    \caption{Observed $Q_{\phi}$ (left) and $U_{\phi}$ (right) images at $K_s$ band. The black arrow indicates the location of the ``dip" in polarized light. Dotted lines have $\pm$45{\degr} slopes and follow transition regions of positive/negative values in $U_{\phi}$. Note the reddish colors in $Q_{\phi}$ which may indicate multiple scattering events (see text for details). 
    }
    \label{fig:QU_multiscat}
\end{figure*}

Figure~\ref{fig:QU_multiscat} shows the observed $Q_{\phi}$ (left) and $U_{\phi}$ (right) images of Sz\,91 at $K_s$ band. For the left panel, we used a different color bar than that of Figure~\ref{fig:obs_img} in order to better visualize changes on disk emission as well as negative values found close to the center of the image (reddish colors). Dotted lines have slopes of $\pm$45{\degr} and, as seen on the right panel, follow transition regions of positive/negative values on the $U_{\phi}$ image. The ``dip" observed on the $Q_{\phi}$ image located at the NW quadrant, and marked by the black arrow, might be related to one of these transition regions on $U_{\phi}$. In fact, at this quadrant is where the $U_{\phi}$ signal is most prominent. We highlight that at these angular positions is where the polarized images are subtracted in order to produce the Stokes parameters (Sect.~\ref{sec:obs}).
Additionally, the negative signal on $Q_{\phi}$ supports the fact that multiple scattering events might be contributing to the total emission \citep{canovas2015c} at least at the disk inner wall; note that the ``dip" is just located at the same radial location and direction (traced by the dotted line) as the NW negative blob. 

All this suggests that the apparent decrease of disk emission in the $Q_{\phi}$ image might not be a real dip on the disk, but rather a hint related to the combined effect of the violation of the strictly azimuthal linear polarization assumption (negative values on $Q_{\phi}$) and/or the data reduction process. Besides, the dip is very faint. We quantified how significant the dip is, compared to the region located at the complementary angle in the southern side, by measuring the emission of the disk along azimuthally distributed apertures and found that it is significant by only 1.2$\sigma$. Also, it is very close to the central star and might be affected by centering effects. Therefore, the veracity of the ``dip" as a real gap or shadow on the disk is unlikely. The fact that the same ``dip" also appears at the same location and with the same direction in the \citet{tsukagoshi2014} observations is certainly intriguing; considering that this data set was taken with a different instrument, at different epochs, and from the northern hemisphere. Nonetheless, given that \citet{tsukagoshi2014} also used the same HWP angles to produce their polarized images may suggest that the ``dip" could be the result of the data reduction process and it is not of astrophysical origin. 
 
\subsection{Polarized phase function}

\begin{figure}
    \centering
    \includegraphics[scale=0.5]{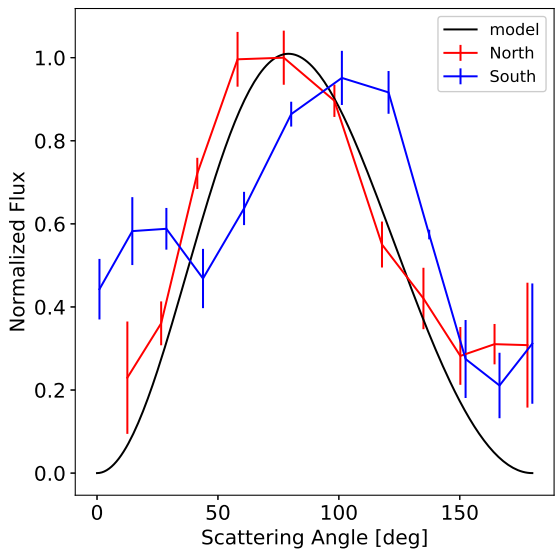}
    \caption{Observed polarized phase function of the northern (red) and southern (blue) sides of Sz\,91. We also plot the polarized phase function of the best fit model for comparison (black). The disk is clearly asymmetric along the minor axis, being brighter at the north side. 
    }
    \label{fig:Phase_function_comparison}
\end{figure}

Figure~\ref{fig:Phase_function_comparison} shows the observed polarized phase function at $K_s$ band of the northern (red) and southern (blue) sides of the disk around Sz\,91. To measure the phase function, we placed adjacent circular apertures along an ellipse that traces the main disk (semi-major axis $a$ of 0.34{\arcsec}, position angle PA of 18.1{\degr} and inclination $i$ of 49.1{\degr}, similarly to our MCFOST modeling). The size of the aperture was fixed to 0.04{\arcsec}. For each aperture we measure the mean flux in the $Q_{\phi}$ image, and the $1\sigma$ uncertainty corresponds to the standard deviation in the aperture. We normalize both the northern and southern sides by the same factor, so that the northern side phase function has a maximum of $1$.
We also plot the phase function of the best fit model (black line) for comparison. As seen on the figure, the disk is clearly asymmetric along the minor axis, with the northern side being brighter than the southern one (something also visible in Figure~\ref{fig:obs_img} and \ref{fig:ALMA_NACO}). Moreover, the disk also exhibits azimuthal asymmetries, for instance the ``dip" discussed in the above section is clearly seen at low scattering angles ($<25{\degr}$).  
Our best model represents relatively well both functions, considering that the MCFOST models used here are centro-symmetric and thus, insensitive to any asymmetric feature (a non-symmetric treatment of the polarized emission is out of the scope of this paper).

The intensity peaks for the phase functions of both the north and south sides are separated by
less than 23{\degr}. We found that this deviation is minimized using a PA of $9{\degr}$ (see Appendix). As stated in Section~\ref{model}, we fixed the PA of the models to the value estimated from the ALMA observations ($18.1{\degr}$), since these are the most sensitive data-set of Sz\,91 to date. Nonetheless, as shown on Figure~\ref{fig:ALMA_NACO} there is a shift between the semi-major axis of the submm ring (given by the white concentric ellipse contours) and the intensity peaks (north-south lobs also in contours) in the ALMA image. In fact, the bright lobes seen on ALMA tend to follow more or less the same (azimuthal) position of the NaCo north-south blobs. This peculiar behavior might explain why a smaller PA produces phase function curves peaking at similar scattering angles at both sides. In any case, the phase function is most sensitive to the properties of the dust (e.g. grain size, porosity) and therefore changes on PA of a few degrees will not affect significantly the results described in Sec~\ref{sec:best_fit} and reported in Table~\ref{tab:best_val}.

The change in PA between the disk semi-major axis, as measured from the ALMA observations, and the intensity peaks of the polarized emission can be due to:
1) a signal-to-noise issue where our observations fail to locate the intensity peaks maximum properly.
2) a projection effect of a flaring disk since ALMA observations probe the disk midplane whereas NaCo observations probe the disk surface layers \citep{stolker2016}. 
3) that the disk that we are detecting with our NaCo observations is slightly warped, suggesting a complex and structured circumstellar disk. Our NaCo observations are not sufficient, and higher SN observations are needed to discriminate between these possible scenarios.

Remarkably, \citet{tsukagoshi2019} also reported an interesting discrepancy regarding the PA of the gaseous disk, going the other way around. From the first-moment map, the authors estimated a PA of $30{\degr}$ for the gaseous disk, which is $12{\degr}$ off compared to the PA estimated for the dust continuum, and $21{\degr}$ compared to the PA that minimizes the phase functions between the northern and southern sides. Nonetheless, this estimate may not be as constrained as for the dust since it may suffer from cloud contamination or by uncertainties in the position of the central star which in turn affects the location of the minimum/maximum velocity in the first-moment map.

Overall, this suggests that the disk around Sz\,91 could be highly structured.

\subsection{Impact of dust properties on scatter polarized emission} \label{sec:dust_prop_pol}

It is expected that grain growth happens via the sticking of small dust grains together. Therefore, a natural consequence is that grains in protoplanetary disks may resemble aggregates build out of small particles (or ``monomers'', see e.g., \citealp{min2016,roy2017,halder2018,tazaki2019}).
Since these aggregates can have different sizes, shapes and with monomers of different compositions, computing realistic optical properties of aggregated particles is a very demanding task. Although exact computations of large aggregates are possible, their use in radiative transfer calculations has been quite limited due to this complexity \citep{min2016,tazaki2019}.
This is why approximate methods, like the DHS which simulates irregular shape, porous aggregates for instance, are usually applied in order to significantly decrease the computational demand without losing the essential information of the aggregates optical properties.    

Our modeling results suggest that the disk probed with our NaCo observations mostly contains very small grains ($<$0.4 {\micron}) in order to reproduce the polarized emission
at $K_s$ and $H$ bands. The grains must be also relatively porous with porosity values lower than 40\%. This implies two possible scenarios: that the emission at the disk surface comes indeed from very small grains, or that large porous aggregates, with radius larger than the wavelength of observation, are present at the upper layers of the disk but the polarized emission we see, is dominated by the small monomers and is insensitive to the global size of the aggregates (one of the conclusions of \citealp{min2016}, see as well \citealp{tazaki2019}). Additional observations of the disk at higher signal-to-noise ratio, in scattered-light (to retrieve the total intensity), and at different wavelengths (e.g., at $J$ band, to measure the color of the disk) may provide new insights in order to discriminate between these possibilities by studying the scattered and/or polarized-light colors and their dependence on size and composition of dust aggregates.     

\section{Conclusions} \label{sec:conclusions}
We present polarized light images at $K_s$ and $H$ band of the $\sim$5 Myr protoplanetary disk around the TTS Sz\,91 taken with VLT/NaCo. We detect a ring-like structure with bright lobes north and south of the
star at both bands. A central cavity is also detected. 
We provide a radiative transfer model that successfully reproduces the main characteristics of the observed polarized emission, and discuss the implications of this study based on the current observational data available for the source. Our main conclusions are as follows: 
\begin{enumerate}
    \item The polarized emission is well reproduced using a disk composed of small (<0.4 {\micron}), porous (<40\%) grains (adopting a distribution of hollow spheres for the scattering theory) with a central cavity of $\sim$45 au in size. Dust grains are most likely in the form of large aggregates and the polarized observations are probably dominated by the small monomers forming the aggregate.
    \item Dynamical clearing by multiple low-mass planets arises as the most likely gap-opening mechanism in Sz\,91. Although, dead zones may account for the presence of gas extended emission inside the dust cavity up to a few au from the central star, a highly depleted gaseous disk beyond the sub-mm ring is also expected. Furthermore, the cavity size in scattered light is expected to have the same size as the sub-mm cavity, which is not the case of Sz\,91. Higher angular resolution observations are needed to confirm the existence of these planets and validate the origin of the disk cavity.
    \item Our $L'$ band mass detection limits put constraints for possible companions of $M_{\rm p}$ < 8 $M_\mathrm{Jup}$ beyond 45au. Within 35au, our sensitivity constraints are poor, and do not rule out the presence of a brown dwarf companion.
    \item The apparent ``dip" observed in the $Q_{\phi}$ image at $K_s$ band is very faint (1.2$\sigma$), and it is most likely the result of the data reduction process and/or contamination by multiple scattering events.
    \item The disk is clearly asymmetric along the minor axis with the north side brighter than the south. We also found a change in PA between the disk semi-major axis, measured from the ALMA observations, and the PA needed to minimize the location of the intensity peaks of the phase functions at the north and south sides of our NaCo polarized observations. This suggests that the disk around Sz\,91 could be highly structured.  
\end{enumerate}

ALMA images with higher resolution and signal-to-noise  capable of resolving the sub-mm ring in the radial direction as well as non-uniform features in the gas around Sz\,91 will undoubtedly help at disentangling between the physical mechanisms behind the origin of the disk cavity. Furthermore, complementary observations in scattered light at different wavelengths using the reference star differential imaging technique, in order to solve the issue of self-subtraction when doing ADI, will also provide new insights about the properties of dust grains at the disk surface layers.

\section*{Acknowledgements}

We thank an anonymous referee for a careful reading of our manuscript and many useful comments.
K.M. acknowledges financial support from FONDECYT-CONICYT project no. 3190859.
K.\,M., J.\,O., M.\,R.\,S., A.\,B., C.\,C., M.\,M., and C.\,P. acknowledge financial support from the ICM (Iniciativa Cient\'ifica Milenio) via the N\'ucleo Milenio de Formaci\'on Planetaria grant.
J.\,O. acknowledges financial support from the Universidad de Valpara\'iso, and from Fondecyt (grant 1180395). C.\,C. acknowledges support from project CONICYT PAI/Concurso Nacional Insercion en la Academia, convocatoria 2015, folio 79150049. M.\,M. acknowledges financial support from the Chinese Academy of Sciences (CAS) through a CAS-CONICYT Postdoctoral Fellowship administered by the CAS South America Center for Astronomy (CASSACA) in Santiago, Chile. L. C. acknowledges financial support from FONDECYT-CONICYT grant no.1171246.

This publication makes use of VOSA, developed under the Spanish Virtual Observatory project supported by the Spanish MINECO through grant AyA2017-84089. 

This research has made use of the NASA/IPAC Infrared Science Archive, which is operated by the Jet Propulsion Laboratory, California Institute of Technology, under contract with the National Aeronautics and Space Administration.

This publication makes use of data products from the Two Micron All Sky Survey, which is a joint project of the University of Massachusetts and the Infrared Processing and Analysis Center/California Institute of Technology, funded by the National Aeronautics and Space Administration and the National Science Foundation.

This paper makes use of the following ALMA data: ADS/JAO.ALMA\#2015.1.01301.S. ALMA is a partnership of ESO (representing its member states), NSF (USA) and NINS (Japan), together with NRC (Canada), MOST and ASIAA (Taiwan), and KASI (Republic of Korea), in cooperation with the Republic of Chile. The Joint ALMA Observatory is operated by ESO, AUI/NRAO and NAOJ.

\appendix
\section{Polarized phase function for a different PA}

In order to find the PA that minimizes the phase functions between the northern and southern sides of the disk around Sz\,91 shown in Fig.~\ref{fig:Phase_function_comparison}, we estimated both phase functions using a range of PA (from 6 to 20{\degr}) and compared the difference between the intensity peaks of both sides in each case. We found that for a PA of 9{\degr} this difference is minimized.

\begin{figure}
    \centering
    \includegraphics[scale=0.5]{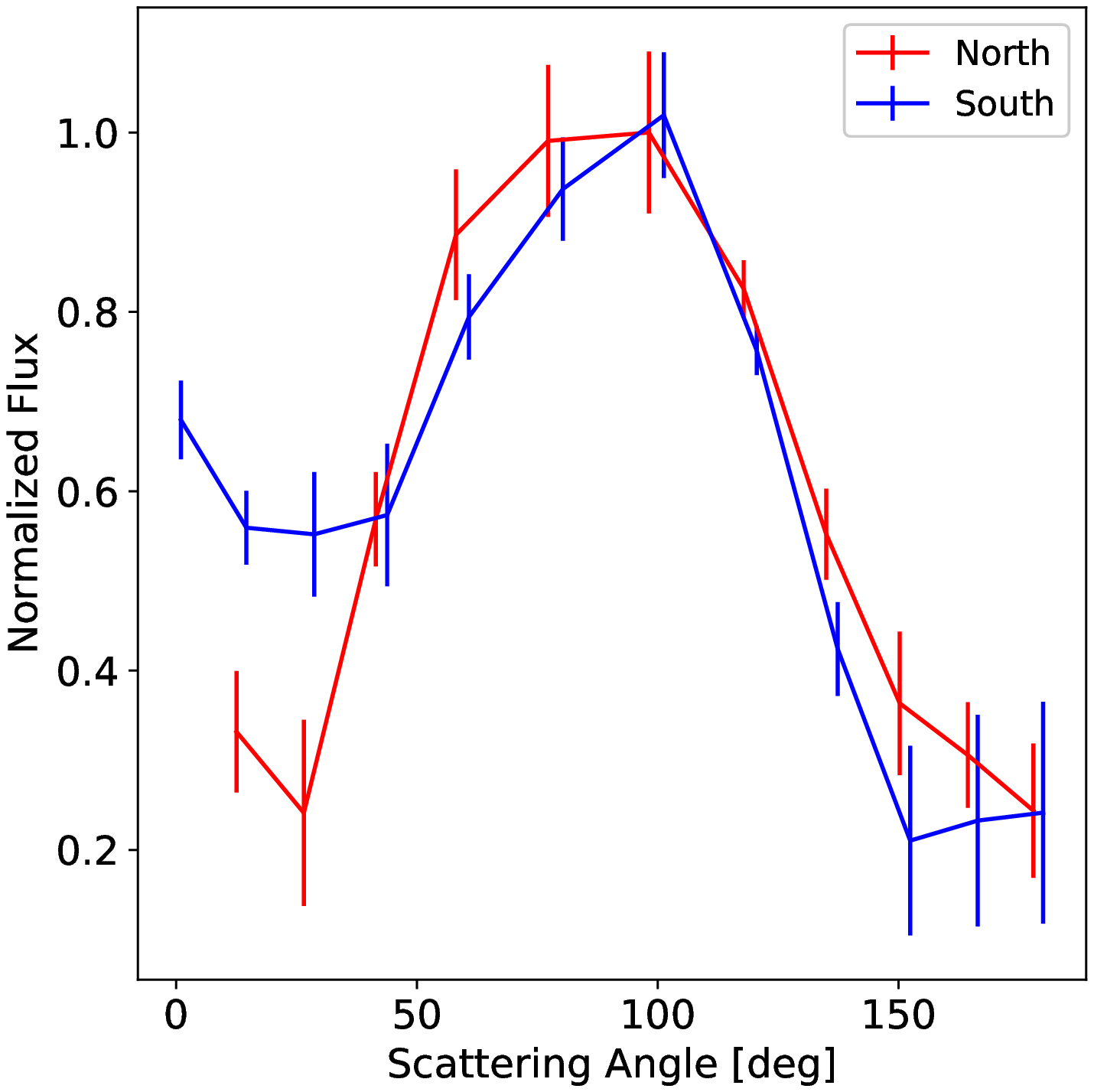}
    \caption{Polarized phase function of the NaCo $K_s$ observations of the northern (red) and southern (blue) sides of the disk around Sz\,91 using a PA = 9{\degr}.
    }
    \label{fig:observed_phase_function}
\end{figure}

Figure~\ref{fig:observed_phase_function} shows the polarized phase function of the NaCo $K_s$ observations of the northern (red) and southern (blue) sides of the disk around Sz\,91 using a PA = 9{\degr}.

\bibliographystyle{mnras}
\bibliography{Mauco}



\bsp	
\label{lastpage}
\end{document}